# Temperature dependent energy levels of electrons on liquid helium

E. Collin[a], W. Bailey, P. Fozooni, P. G. Frayne, P. Glasson, K. Harrabi[b] and M. J. Lea[c]
*Department of Physics, Royal Holloway, University of London, Egham, Surrey, TW20 0EX, England.*
(12 November 2017)

We present measurements of the resonant microwave absorption between the Rydberg energy levels of surface state electrons on the surface of superfluid liquid helium, in the frequency range 165 – 220 GHz. The resonant frequency was temperature dependent. The experiments are in agreement with recent theoretical calculations of the renormalisation of the electron energy levels due to zero-point and thermal ripplons. The temperature-dependent contribution to the linewidth $\gamma(T)$ for excitation to the first excited state at 189.6 GHz is compared with other measurements and theoretical predictions.

## I. INTRODUCTION

Electrons above the surface of liquid helium [1] are a fundamental two-dimensional conducting system. The electrons move in a potential well formed by a high potential barrier at the liquid surface and the attractive Coulomb potential of a weak positive image charge in the helium $U(z) = -Ze^2/4\pi\varepsilon_0 z$, with $Z = (\varepsilon - 1)/4(\varepsilon + 1) = 0.007$. This gives a hydrogen-like spectrum, with energy levels $E_m \approx -R_e/m^2$, where the effective Rydberg energy $R_e = m_e e^4 Z^2/8h^2\varepsilon_0^2 \approx 0.67$ meV for liquid $^4$He, $m$ ($\geq 1$) is the quantum number and ε is the dielectric constant of the helium.

For 2-D free electrons, the Rydberg levels give a series of sub-bands. Microwave transitions between the sub-bands were first observed by Grimes *et al.* [2] above 1.2 K for transitions from the ground state to the exited states $m = 2, 3$. Because of the asymmetry of the wave-functions, these energy levels can be tuned with a vertical electric field $E_z$, in a linear Stark effect. The resonant frequency $f_{21} = (E_2 - E_1)/h$ increases from 125.9 GHz in zero holding field up to 220 GHz for $E_z = 17.5$ kV/m. Lambert and Richards [3] extended the frequency range up to 765 GHz using a far-infrared laser.

Edel'man [4] and later Volodin and Edel'man [5] indirectly probed the Rydberg states, on both liquid $^4$He and $^3$He, by measuring the photoconductivity when the electrons were excited by incident microwaves.

Collin et al. [6] extended the measurements on liquid $^4$He down to the ripplon scattering region below 0.7 K**.** They also investigated the absorption of microwaves at high powers and low temperatures in the non-linear region, but underestimated the effects of electron heating. We now present an extended analysis of these experiments and, in particular, unpublished results on the temperature dependence of the resonance frequency $f_{21}(T)$ in comparison with recent theory.

Isshiki et al. [7] measured the microwave absorption for electrons on liquid $^3$He from 0.01 to 1 K, with some experiments on liquid $^4$He. Further microwave experiments have explored non-linear effects in electrons on helium [8, 9, 10, 11, 12].

The theory was given by Ando [13] for both the gas-atom and ripplon scattering regimes. This underestimates the linewidth $\gamma(T)$ for vapor-atom scattering by a factor of 1.6 on $^4$He and 2.1 on $^3$He.

The binding energy of a surface-state electron to liquid helium is a conceptually simple problem, yet remains controversial. Grimes *et al.* [2] modeled this using an instructive empirical potential

$$U(z) = \frac{-Ze^2}{4\pi\varepsilon_0(z+b)} \quad \text{for } z > 0 \qquad (1)$$
$$= V_0 \quad \text{for } z \leq 0$$

The effect of $V_0$ is to lower the ground state energy, as the electron wavefunction penetrates into the helium with a decay length $\delta = 0.195$ nm for $V_0 = 1.0$ eV, and hence to increase the transition frequency $f_{21}$. A diffuse helium surface increases the ground state energy. The fractional correction to $f_{21}$ is $14(\delta - b)/3a_B$ where $a_B$ is the effective Bohr radius (7.6 nm for electrons on $^4$He). Grimes *et al.* [2] fitted the data to $V_0 = 1.0$ eV and $b = 0.104$ nm while Volodin and Edel'man [5] found $V_0 = 1.3$ eV and $b = 0.125$ nm for $^3$He.

There have been many phenomenological models [2, 5, 14, 15, 16, 17] for the electron energy spectrum. Most recently, Degani *et al* [18] estimated the temperature dependence, predicting a large, almost linear, decrease with temperature for $E_z = 0$.

But these models treat the surface profile and hence the potential seen by the electrons, as static, though it can be temperature dependent. Zero-point and thermal ripplons on the helium surface [19] contribute to the diffuse surface or interfacial profile [20, 21]. The electrons also react adiabatically to ripplons. Konstantinov *et al.* [22] show this renormalises the electron energy spectrum at zero temperature and gives a non-linear temperature dependence of the resonant frequencies down to the lowest temperatures. This ripplon-induced frequency shift, arising from the combination of a high interface barrier and a randomly distorted surface, including both zero-point and thermal ripplons, is subtle and involves both 1-ripplon and 2-ripplon interactions. This effect is a condensed matter analogue of the Lamb shift in the H atom, in which the electron wavefunction is 'blurred' by interactions with zero-point photons in free space, giving a small shift in the binding energy.

---

[a] Present address: Institut Néel, CNRS & Université Grenoble Alpes, 25 rue des Martyrs, BP166, F-38042, Grenoble Cedex 9, France.
[b] Present address: Physics Department and Center of Research Excellence in Renewable Energy, King Fahd University of Petroleum and Minerals, 31261 Dhahran, Saudi Arabia.
[c] Corresponding author: m.lea@rhul.ac.uk.

The theory contains diverging terms from short-wavelength ripplons but these effectively cancel, leaving only a small increase in the energy levels with temperature $\Delta E_m(T)$. The corresponding temperature dependent shift of the resonant frequency $f_{21}$ is

$$\Delta f_{21}(T) = [(\Delta E_2(T) - \Delta E_1(T)]/h < 0 \qquad (2)$$

This paper describes measurements of the resonant microwave absorption of surface state electrons on liquid $^4$He to temperatures below 1 K. Measurements of the linewidth $\gamma_{21}(T)$ are presented and compared with other experiments and the theory of Ando [13]. The temperature dependence of the resonant frequency $\Delta f_{21}(T)$ is in good agreement with Konstantinov *et al.* [22] below 1 K.

## II. MICROWAVE ABSORPTION

### A. Lorentzian lineshape

The microwave coupling between the electronic wavefunctions on helium is very strong with a dipole transition length $z_{21}$ between the ground state $\psi_1(z)$ and the first excited state $\psi_2(z)$

$$z_{21} = \int \psi_1^*(z) z \psi_2(z) \mathrm{d}z \qquad (3)$$

which increases from $z_{21} = 0.59\, a_B = 4.25$ nm for $E_z = 0$ to $z_{21} = 5.14$ nm for $E_z = 10.6$ kV/m, corresponding to $f_{21} = 190$ GHz. In the low power limit, the integrated microwave absorption, normalized to the peak, is expected to have a Lorentzian lineshape [13]

$$L(\delta f, \gamma) = \frac{\gamma^2}{\delta f^2 + \gamma^2} \qquad (4)$$

where $\delta f = f - f_{21}$. The microwave linewidth $\gamma$ for electrons on helium has been calculated by Ando [13]. Two contributions can be distinguished, $\gamma = \gamma_1 + \gamma_2$. The first contribution $\gamma_1$ comes from the inter-sub-band transitions from the excited state to the ground state with a transition rate $1/\tau$, with $\gamma_1 = 1/2\tau$ for gas-atom scattering, with

$$\frac{1}{\tau} = \frac{\sigma_a N_g(T)}{4 m_e a_B} \int_0^\infty \psi_1^2(z) \psi_2^2(z) \mathrm{d}z \qquad (5)$$

The second term corresponds to decoherence, from intra-sub-band scattering with no transition in the Rydberg state, with $\gamma_2 = 1/\tau_2$. The scattering time $\tau_2$ for gas-atom scattering (equivalent to $\delta$–function scattering) is given by[13]

$$\frac{1}{\tau_2} = \frac{\sigma_a N_g(T)}{4 m_e a_B} \int_0^\infty (\psi_1^2(z) - \psi_2^2(z))^2 \mathrm{d}z \qquad (6)$$

where $\sigma_a = 4.98 \times 10^{-20}$ m$^2$ is the helium atom scattering cross-section and the vapor atom number density $N_g(T) = (2\pi m_4 kT/h^2)^{3/2} \exp(-7.17/T)$ where $m_4$ is the mass of a $^4$He atom. The total temperature dependent linewidth is given by

$$\gamma(T) = AT + BN_g(T) \qquad (7)$$

where the first term is due to ripplon scattering. Coefficients $A$ and $B$ depend weakly on the holding field $E_z$ [13].

Experimental linewidths $\gamma(T)$ in this paper are half-widths at half-height (HWHH), given in frequency units of MHz, as converted from the measured voltage sweep in $V_z$ and the slope of Fig. 3 with $\beta = 2.44$ GHz/volt.

### B. Microwave cell

Microwave experiments were performed at frequencies between 165 and 260 GHz, though most were done at 189.6 GHz. Further information about the microwave system is given in Supplemental Material.

The microwave cell is shown schematically in Fig. 1. The electrons were held above the liquid helium between capacitor plates $D = 2.075$ mm apart, which formed a flat cylindrical cavity, 56 mm in diameter. The upper and lower electrodes consisted of a central circular disk, surrounded by an annular ring. The lower ring was divided into four segments. The electrons were produced by thermionic emission from a pulsed filament in the upper cell.

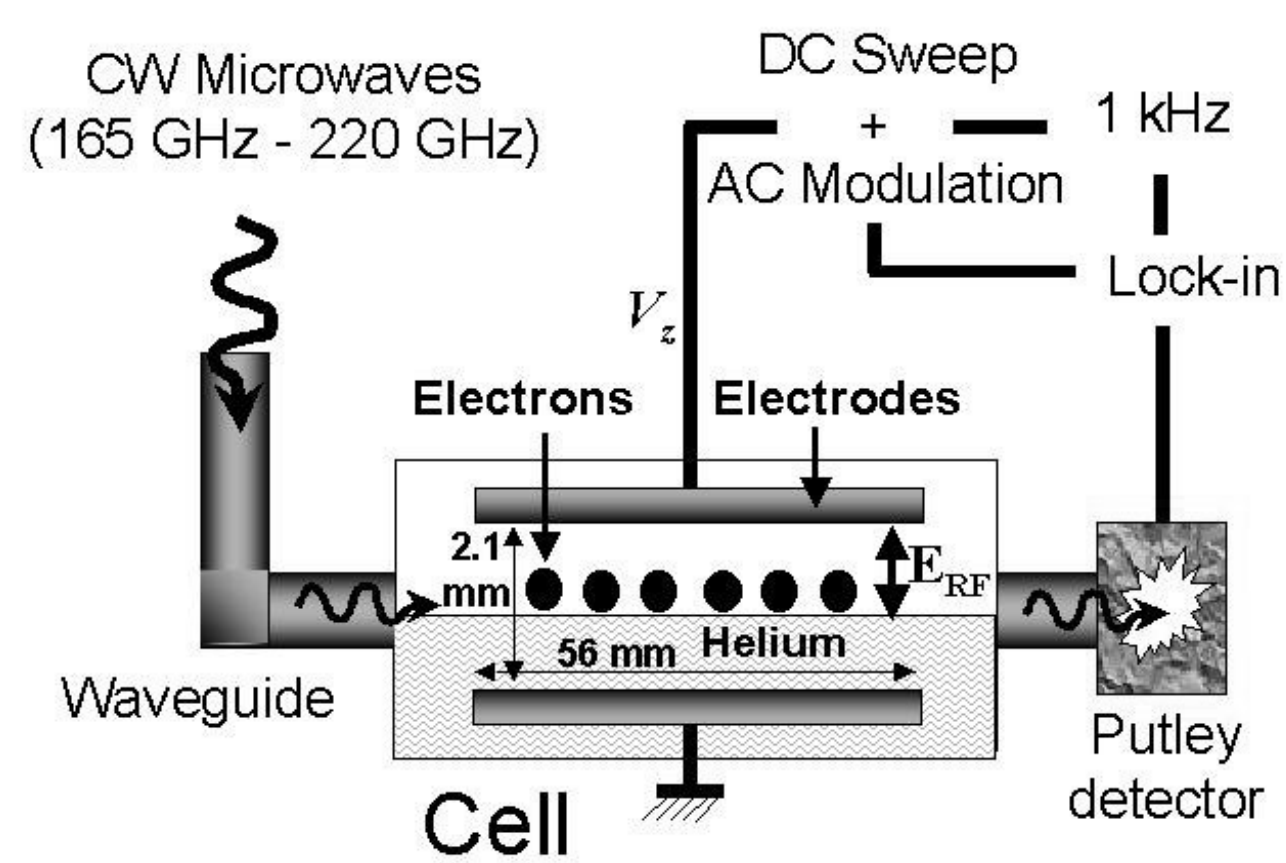

FIG. 1. Schematic diagram of the microwave cell. The cell was half-filled with liquid helium as a substrate for the surface-state electrons. Vertically polarized microwaves were transmitted through the cell and detected by an InSb Putley detector. The vertical electric field $E_z$ was swept through the microwave resonance. The field was a.c. modulated and the modulated microwave signal was detected using a lock-in amplifier.

The vertical separation of the cell was determined by two factors. First, the capillary length of liquid helium, $l_c = (\eta/\rho g)^{0.5} = 0.50$ mm, where η is the surface tension and ρ is the density of the liquid helium. When the cell is almost full, capillary rise will produce a "bubble" in the center of the cell of thickness $2l_c$ (like the bubble in a spirit level) and the cell would then fill from the circumference. Hence it is not possible to have a stable and flat helium surface across the center of the cell with $D < 4l_c = 2.0$ mm, and this is the minimum height of the cell. The second reason is that the propagating microwave mode required is $p = 2$, where $p$ is the number of standing half-wavelengths along the $z$-axis. For this mode the $z$-axis component of the microwave field is a maximum on the surfaces of the upper and lower electrodes and in the middle of the cell, at the helium surface where the electrons are distributed. The low frequency cut-off for the $p = 2$ mode is $f_c \approx c/D = 150$ GHz, well below the operating frequency. Microwave transmission experiments clearly showed the onset of these modes in test cavities, as the frequency was increased above cut-off. The central part of the cell forms a microwave pillbox cavity, with a complicated series of TM mode resonances as the microwave frequency was swept. The in-plane wavelength at 190 GHz is about 2.3 mm and the radial intensity distribution will form a pattern of nodes and antinodes from radial and azimuthal standing waves.

Continuous wave (cw) microwaves were used with two a.c. modulation schemes. Sine-wave modulation, typically 10 mV rms at 0.5 to 10 kHz, of $V_z$ was used to measure the differential absorption $\alpha' = d\alpha/dV_z$ with a lock-in amplifier and integrated to obtain the absorption lineshape $\alpha(V_z)$. Alternatively large amplitude square-wave modulation, typically ± 1.5 V, much bigger than the linewidth, gave the microwave absorption line directly, without numerical integration. Both schemes gave identical results at low power levels. The modulation was imposed by applying equal in-phase and anti-phase voltages to the upper and lower electrodes.

### C. Electric holding field

For each experiment, fixed frequency microwaves were used. The frequency $f_{21}(T)$ was tuned through resonance using the vertical holding field $E_z = E_{z1} + E_{z2}$, which is proportional to the voltage difference $\Delta V = V_{upper} - V_{lower}$ across the capacitor plates

$$E_{z1} = \frac{-\Delta V}{(D - d + d/\varepsilon)} \tag{8a}$$

where $D$ is the vertical separation of the cell electrodes and $d$ is the helium depth. The field was swept by symmetrically sweeping the potentials on the upper and lower electrodes, to maintain a constant potential on the electron sheet. To confine and hold surface electrons $\Delta V = V_{upper} - V_{lower} < 0$. However in the experimental plots and discussion we define $V_z = V_{lower} - V_{upper} > 0$ as the pressing voltage, giving an upwards pressing field $E_z$ for the negatively charged electrons.

The field also depends on the electron density $n$. The density term $E_{z2}$ is more subtle. For a uniform 2-dimensional electron sheet in a grounded parallel plate capacitor partly filled with dielectric, the induced charges produce a vertical field above the dielectric

$$E_{z2} = -\frac{ne}{2\varepsilon_0}\frac{(D - d - d/\varepsilon)}{(D - d + d/\varepsilon)} \tag{8b}$$

But this expression assumes that the polarisation charges $ne(\varepsilon-1)/(\varepsilon+1)$ in the upper surface of the liquid helium are uniformly distributed. For electrons which are much closer to the helium surface than their in-plane spacing ($z << n^{-1/2}$), individual image charges give rise to the $1/z$ binding potential for each electron and should not be included in $E_z$. Hence the constant electric field seen by individual electrons close to the helium surface is then given by [24]

$$E_z = \frac{-\Delta V}{(D - d + d/\varepsilon)} + \frac{ne}{\varepsilon_0(\varepsilon+1)}\frac{(D - 2d)}{(D - d + d/\varepsilon)} \tag{8c}$$

The field $E_z$ is independent of $n$ only if the helium surface is at the geometrical center of the cell, $d = D/2$. For $z >> n^{-1/2}$ the density dependent field is given by Eq. (8b).

### D. Electrons on liquid helium

The cell was filled with ultra-pure $^4$He [23]. The helium level in the cell was monitored during filling by measuring the capacitance between the upper and lower electrodes. The cell was leveled by charging the helium with electrons from a pulsed filament, and then driving the upper central electrode with a 10 kHz a.c. voltage and measuring the a.c. current to the 4 segmented electrodes on the lower electrode. The cryostat was tilted using air-mounts to balance these 4 currents, thus leveling the helium in the cell. The tilt of the cryostat was continuously monitored during the experiments using variable resistance displacement devices on the cryostat top-plate.

The electron density $n$ was measured by using a Corbino electrode geometry to measure the magnetoconductivity [24]. A 10 kHz a.c. modulation was applied to the central electrode and the a.c. current flowing to the outer ring via the electrons was measured. A resistive phase shift $\varphi \propto 1/\sigma_{xx}$ is observed and hence $\sigma_{xx}$ can be determined. In zero magnetic field $\sigma_0 = ne\mu$ where μ is the electron mobility. In a small perpendicular magnetic field $B$ (< 1 Tesla), the Drude law holds [24], with $\sigma_0/\sigma_{xx}(B) = 1 + \mu^2 B^2$. Magnetoconductivity measurements enabled both $\sigma_0 = ne\mu$ and $\mu B$ to be found independently allowing μ and $n$ to be determined. The values of μ were in good agreement with previous experiments [24]. The electron density profile was uniform out to a radius of about 27 mm.

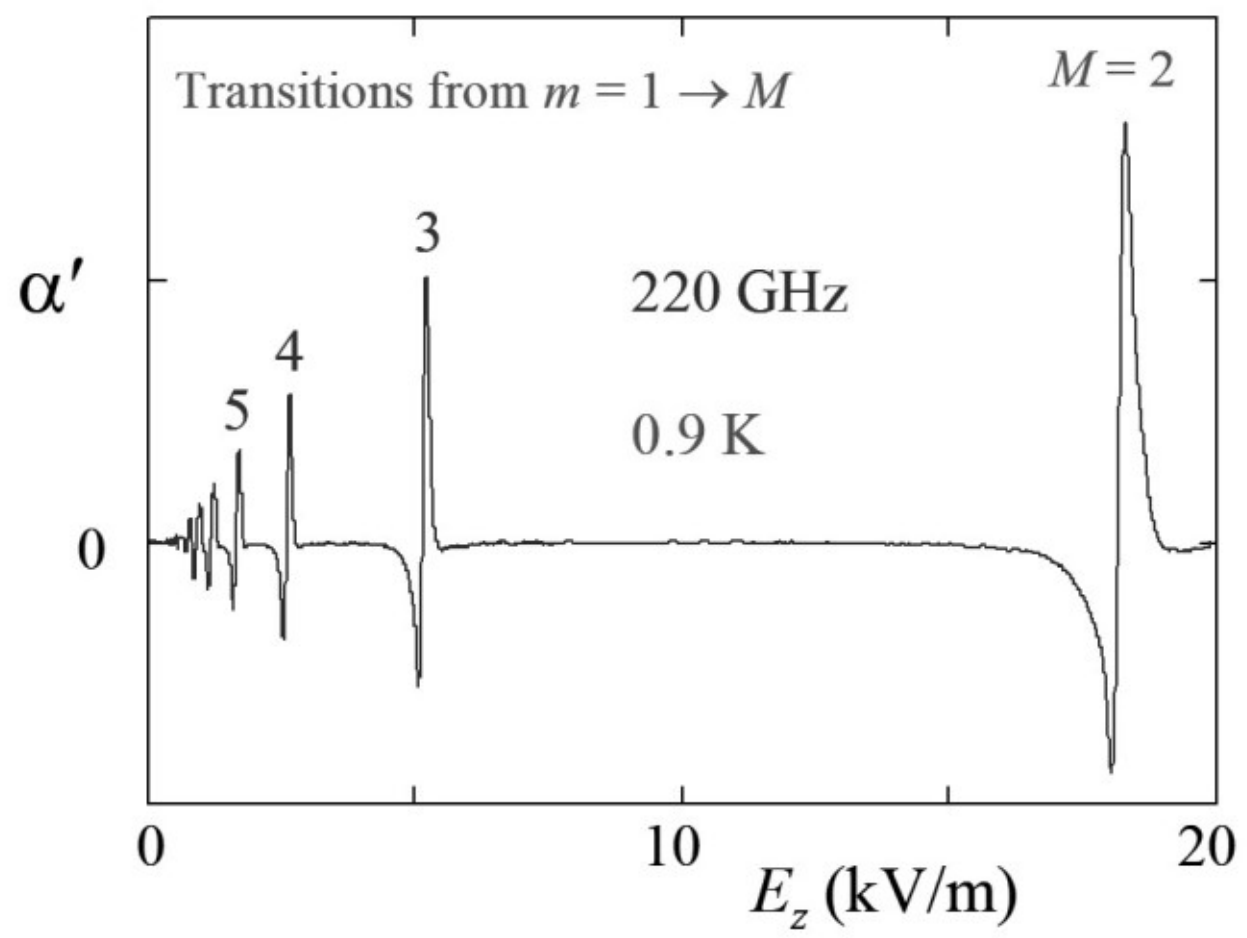

FIG. 2. The differential microwave absorption α′ at 220 GHz as the vertical holding field $E_z$ was swept, showing resonant transitions from the ground state (sub-band), $m = 1$, to higher Rydberg states (sub-bands) $m = M$ (with $M$ up to 12 observed).

## III. RESULTS AND ANALYSIS

### A. Resonant frequency

As the vertical holding field $E_z$ was swept over a wide range at a fixed microwave frequency, a series of absorption peaks was observed as shown in Fig. 2 at a frequency of 220 GHz. These correspond to transitions from the ground state (sub-band), $m = 1$, to higher Rydberg states (sub-bands) $m = M$, as the Coulomb potential well changes with $E_z$. In these experiments we are primarily interested in the transition from the ground state to the first excited state, $M = 2$. This resonance was measured over the frequency range 165 to 220 GHz as shown in Fig. 3, where $f_{21}$ is plotted versus $E_z$ at 1.3 K, confirming the original data from Grimes and Adams. The solid line shows their model calculations.

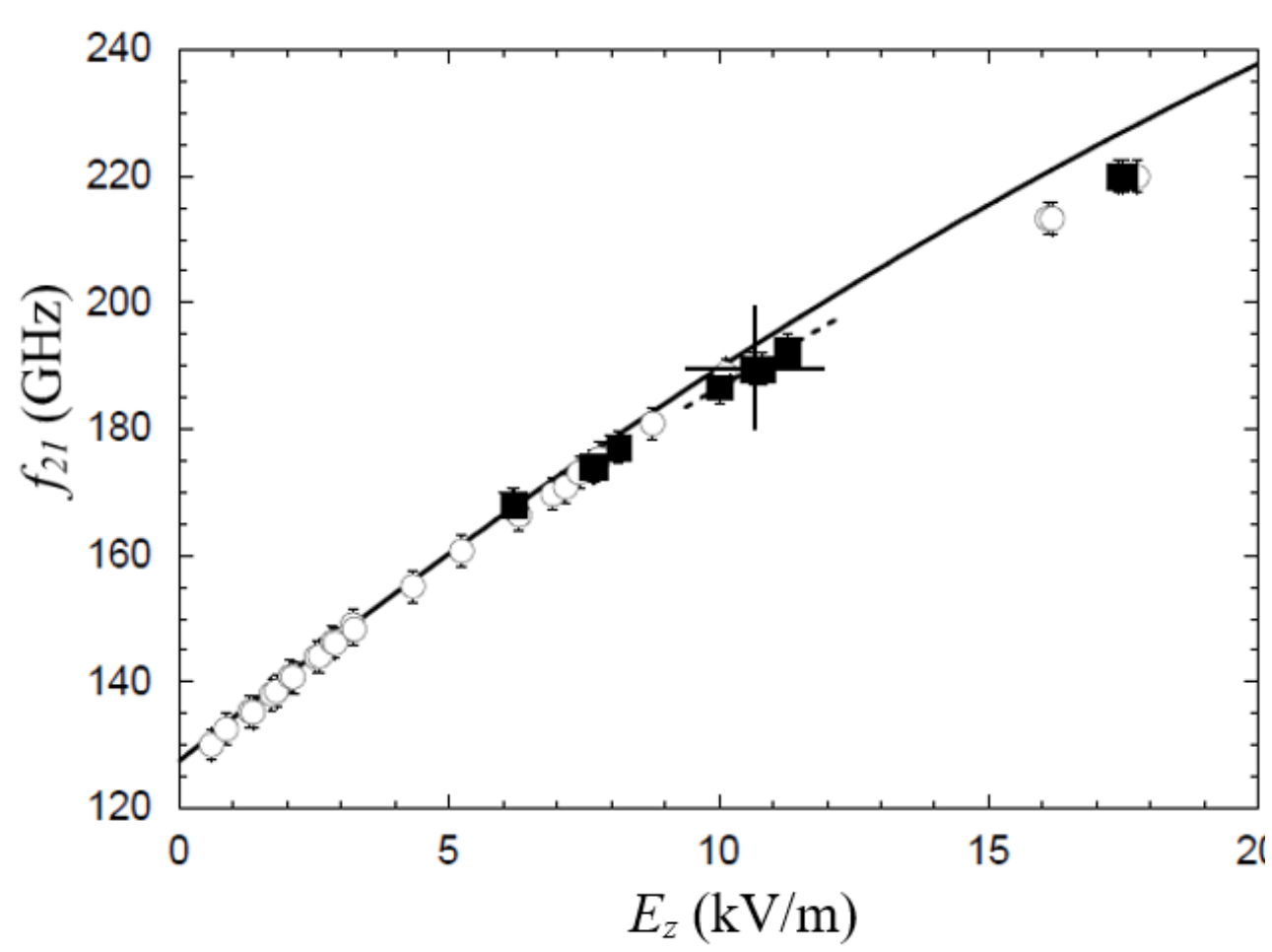

FIG. 3. Resonant frequency $f_{21}$ versus $E_z$ at 1 K, for low microwave powers (■). Measurements from Grimes *et al* [2]. at 1.2 K are shown (○), while the solid line shows their theoretical calculations. The cross indicates the frequency 189.6 GHz for our temperature dependent measurements. The dashed line is the slope of the experimental curve at this point.

Most of the measurements were made at a fixed applied microwave frequency of 189.6 GHz, where a local maximum in the cell response was observed. The slope of the graph at 189.6 GHz corresponds to 2.44 GHz/volt for a sweep in $V_z$.

### B. Absorption lineshape

The measured differential absorption α′ was integrated numerically to give the absorption lineshape $\alpha(V_z, T)$. Above 1 K it is close to a Lorentzian, but at low temperatures and low microwave power it can have a complex structure as shown in Fig. 4 at 0.304 K and in Fig. 5 at 0.304, 0.855 and 1.004 K. The data in Fig. 5 is plotted against the pressing voltage $V_z$ while in Fig. 4 it is plotted against the fractional change $\Delta E^*$ in the vertical electric field $E_z$. as $V_z$ is swept through the resonance ($\Delta E^* = 0$). Two resonance peaks are seen very close together which reflect the electrode geometry.

Both peaks at 0.304 K can be precisely fitted by a Lorentzian, linewidth (dotted line) $\gamma = 27 \pm 2$ MHz $\equiv 11$ mV $= 5.04 \times 10^{-4}$ in the fractional field, broadened by the small a.c. modulation of 10 mV rms ($6.4 \times 10^{-4}$ in the fractional field). The offset second peak with the same parameters is followed by a long tail.

Sloggett et al.[25] numerically analysed the vertical electric field distribution in circular disc capacitors, defining δ as the fractional change in field compared to infinite electrodes with the same spacing and voltage. The field is maximum at the center with a deviation $\delta_0 = -2.57 \exp(-6.079\ R/D)$. It decreases exponentially towards the edges as $\delta(r)$ = -exp(-(1+2π($R$-$r$)/$D$), reaching $\delta \approx -0.01$ at 1 mm from the outer edge of the capacitor The first peak comes from electrons between the 22 mm radius central electrodes ($\delta_0 \approx -2.3 \times 10^{-29}$) with 59% of the electrons. There is a 0.5 mm gap between the central and ring electrodes. This gives a small decrease in $E_z$ over the gap region. The second peak comes from electrons (26% of the total) between the ring electrodes with a central fractional field offset of $\delta_0$ = -0.0028, in good agreement with Sloggett for an electrode size of 4.5 mm. The tail comes from the remaining 15% of electrons under the gap and round the edge where the field reaches $\delta \approx -0.01$ at the edge of the electron sheet, as seen in Fig. 4. The rms inhomogeneity in $E_z$ across the electron sheet is 0.2%.

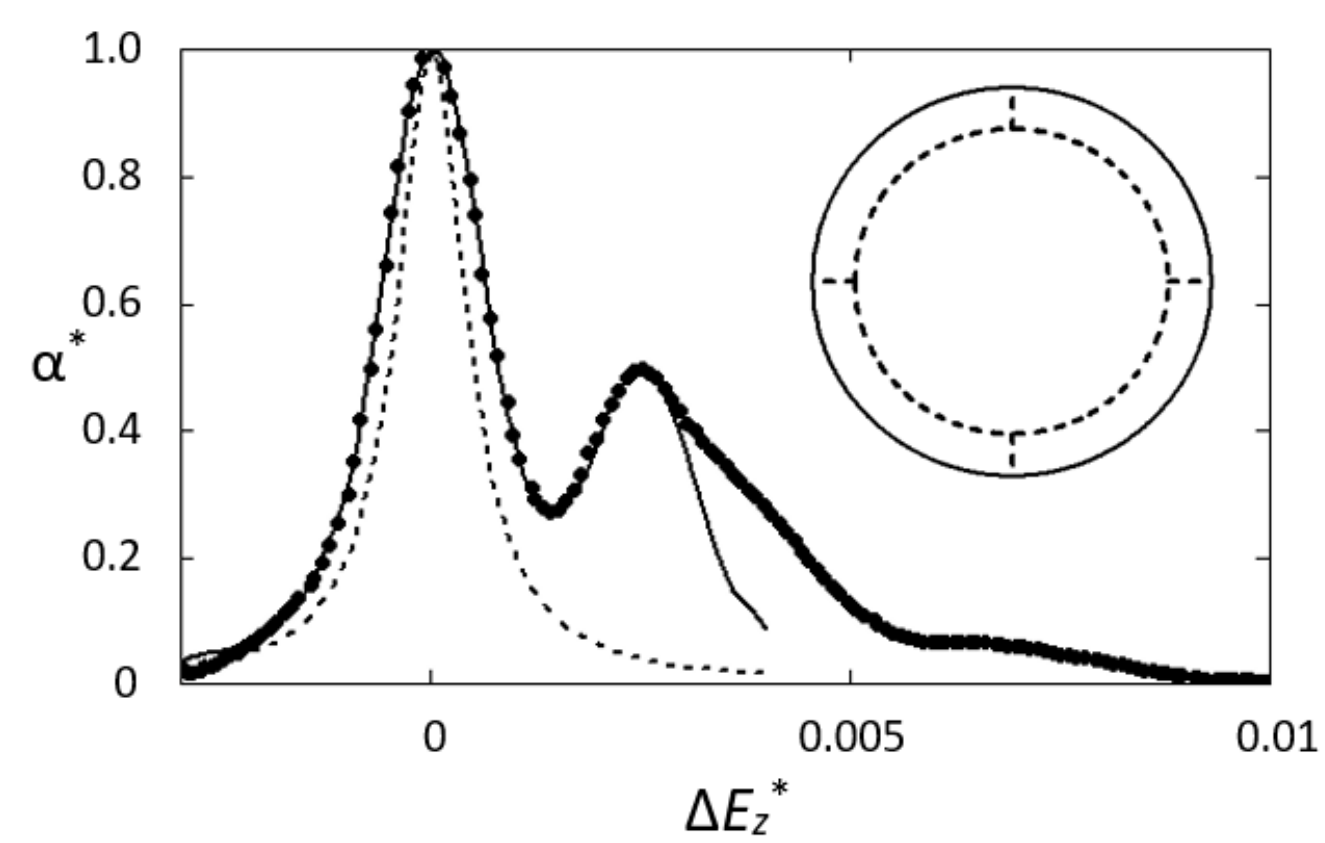

FIG. 4. The measured absorption lineshape α at 189.6 GHz at 0.304 K, normalised to the first maximum and plotted against the fractional change in vertical electric field as the voltage is swept. The inset shows the electrode geometry, outer radius 28 mm, with the 4 ring segments in the lower electrode, inner radius 22 m. The solid line shows a Lorentzian, $\gamma = 5.04 \times 10^{-4}$ (dashed line), broadened by a field modulation of 10 mV rms. The second peak has a relative magnitude 0.46 and a field offset $\delta = -0.0028$.

However the linewidth of 27 MHz at 0.3 K can only be regarded as an upper limit for any residual or zero-point linewidth. An rms variation of only 1 μm in the electrode separation across the 56 mm diameter would give a residual linewidth of some 38 MHz ($7 \times 10^{-4}$ in the fractional field). Other small effects on the lineshape at this level could include non-parallelism of the electrode plates, a tilt in the helium surface and standing waves in the microwave intensity.

We used the experimental lineshape at $T_0$ = 0.3 K as a reference $\alpha_0(V_z)$ for analysis at higher temperatures, convolving it with a temperature dependent Lorentzian to obtain $\gamma(T)$. The precision (< 1 MHz) of these fits was excellent (solid lines in Fig. 5), clearly showing both peaks broadening and shifting together.

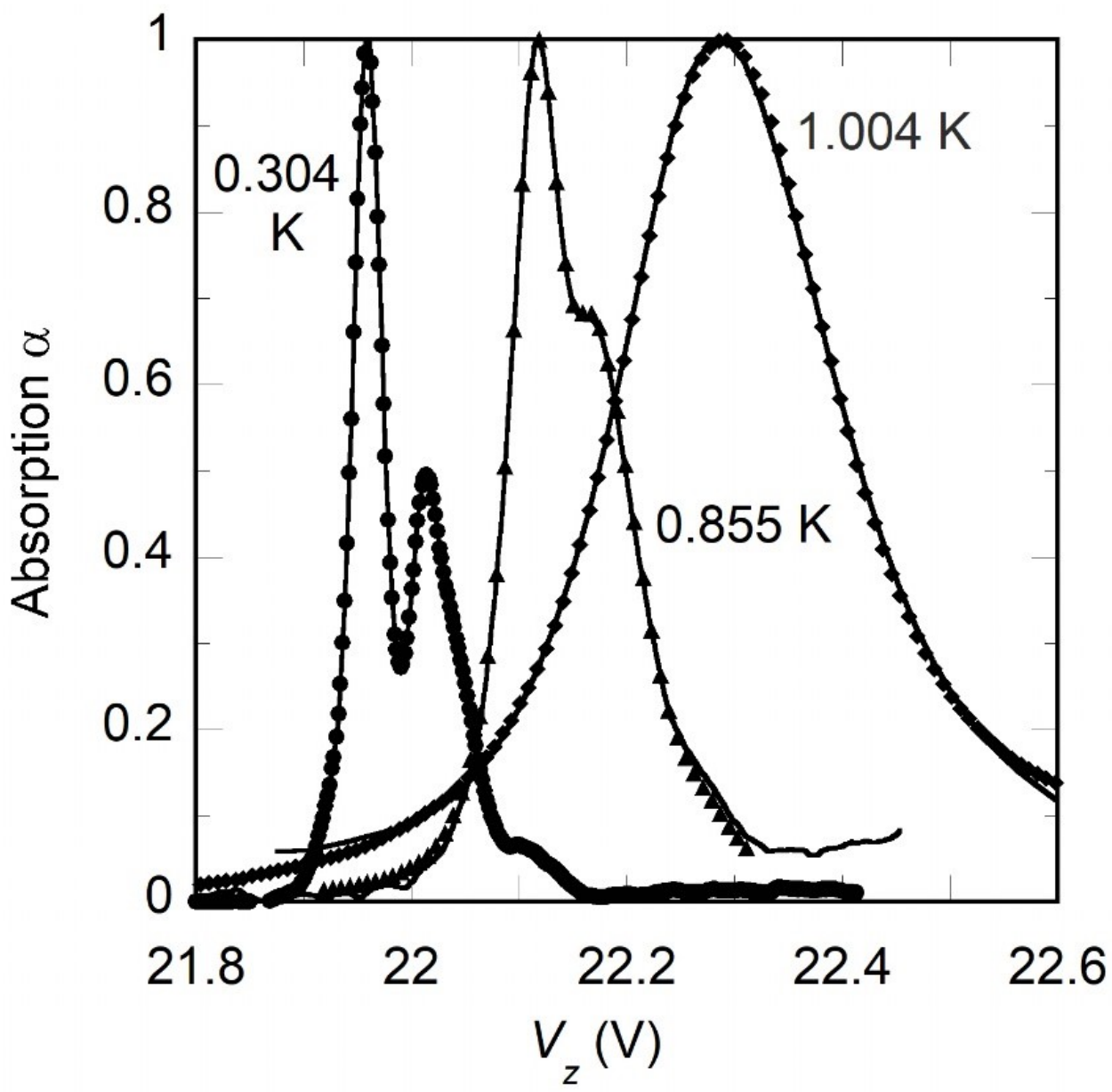

FIG. 5. Temperature dependent lineshapes. The inhomogeneously broadened lineshape is shown at 0.304 K (●). The lineshape broadens as the temperature increases as shown at 0.855 K (■) and 1.004 K (♦). The lines through the points show the convolution of the 0.304 K lineshape with a Lorentzian lineshape with $\gamma$ = 49 MHz at 0.855 K and 288 MHz at 1.004 K. The lineshapes have been normalized to their peak values. The position of the line is temperature dependent, as discussed in the text.

In the low power limit, the area of the line should be proportional to the total number of electrons and the incident microwave power. Experimentally, the area was indeed independent of $T$ at low powers. The area was used to estimate the electron density for each set of data, relative to reference data sets in which the density was measured directly from the magnetoconductivity. The typical electron density for the linewidth data presented here was $1.7 \times 10^{11}$ m$^{-2}$, though densities up to $6.5 \times 10^{11}$ m$^{-2}$ were used. No dependence of the linewidth on the electron density was observed. At higher powers and temperatures above about 1.5 K, we gradually started to lose electrons.

At higher microwave powers, we observed power broadening, absorption saturation, non-linear effects and hysteresis. The Lorentzian linewidth becomes power dependent $\gamma_P^2 = \gamma^2 + \gamma\tau\Omega^2 = \gamma^2 + bP$, where the incident microwave power $P \propto E_{RF}^2 \propto \Omega^2$ where $\Omega = eE_{RF}z_{12}/\hbar$ is the Rabi frequency, $E_{RF}$ is the microwave field amplitude and $1/\tau$ is the spontaneous decay rate from the excited state. This leads to absorption saturation and power broadening. At low powers, we recover a Lorentzian lineshape, Eq.(4).

### C. Temperature dependent linewidth

The lineshape was analysed to obtain $\Delta\gamma(T) = \gamma(T) - \gamma(0.3)$. From Eq. (7) the theoretical temperature dependent linewidth $\gamma(T) = AT + CN_g(T)/N_g(1)$ where the first term is due to ripplon scattering and the second to vapor atom scattering. The experimental reference value of $\gamma(0.3)$ was determined by a linear fit of the temperature dependence of $\Delta\gamma(T)$ below 0.7 K and extrapolating to zero temperature, following Eq. (7). The resultant experimental values of $\gamma(T)$ are shown in Fig. 6. The solid line shows numerical calculations of $\gamma(T)$ from Ando's theory [13], with $A$ = 12.8 MHz/K and $C$ = 109 MHz. The transition from gas-atom to ripplon scattering is clearly seen at about 0.7 K. The fit below 0.7 K is good, though with only a few data points.

The solid line shows Ando's theoretical calculations, which lie below the experimental results. The dashed line shows the vapor-atom scattering scaled up by a factor of 1.6 in agreement with Isshiki *et al*. [7].

### D. Temperature dependent resonant frequency

We also found that the resonant frequency itself was unexpectedly, and rather strongly, temperature dependent. The experiments were done at a fixed microwave frequency of 189.6 GHz for a range of electron densities $n$. The resonant absorption frequency of the electrons was tuned using the Stark effect by sweeping the top-plate voltage $V_z$ and hence the vertical electric field $E_z$.

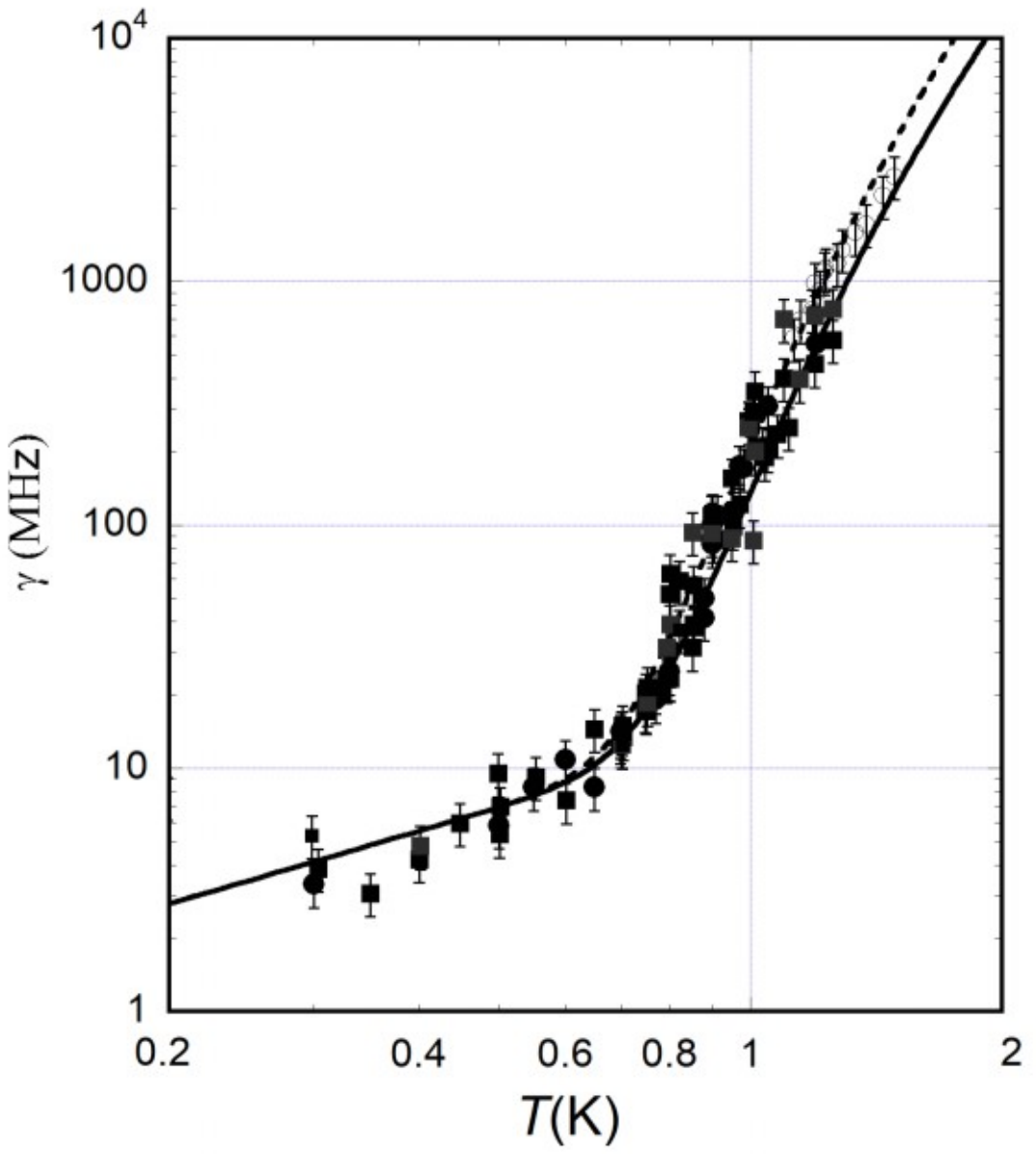

FIG. 6. Temperature-dependent contribution to the Lorentzian linewidth γ($T$) versus temperature at 189.6 GHz. Measurements of the lineshape height and width (■, ●) were used to obtain $\Delta\gamma(T) = \gamma(T) - \gamma(0.3)$, relative to the value at 0.3 K. Data from Grimes et al [2], scaled for the different pressing field, is shown (○). The solid line is the theory by Ando [13], while the dashed line shows Ando's theory with the vapor-atom scattering contribution scaled by a factor of 1.6.

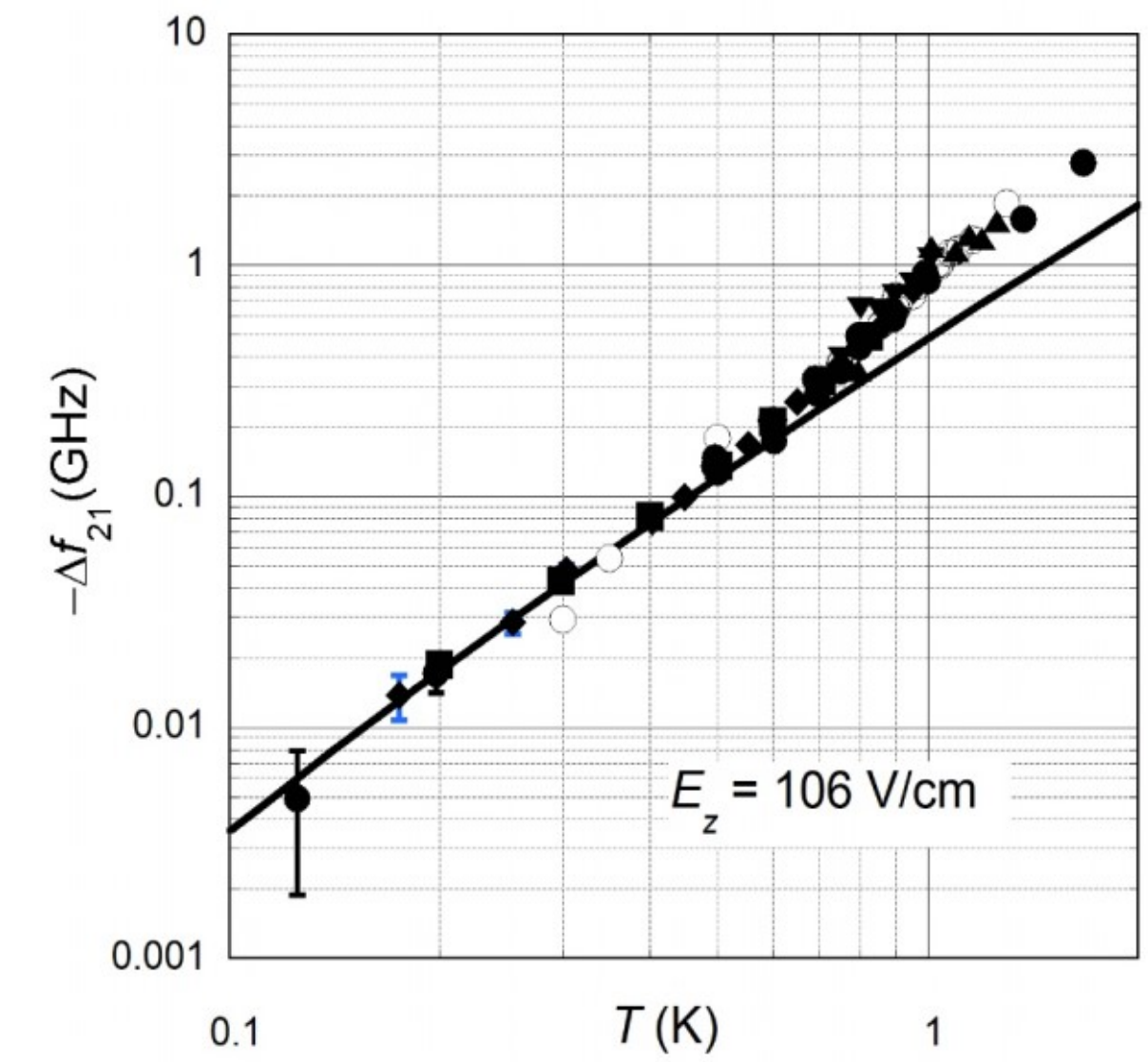

FIG. 7. The temperature dependent resonance shift $\Delta f_{21}(T) = f_{21}(T) - f_{21}(0)$ at 189.6 GHz. The resonance position moved to higher values of $V_z$ at higher temperatures. The voltage shift was converted to a frequency shift using the factor 2.44 GHz/V. The electron densities were 0.67 (▲), 1.0 (▼), 1.5 (♦), 1.7 (○) and 2.4 (■) × $10^{11}$ m$^{-2}$. Other data points (●) were for various electron densities. The solid line shows the theoretical calculations of Konstantinov *et al.* [22] due to ripplon renormalisation.

Following the lineshape analysis, the voltage $V_{21}(T)$ at resonance was obtained from 0.1 to 1.5 K, for a range of electron densities $n$. As the temperature increased, the voltage increased as shown in Fig. 4. This corresponds to a decrease in the resonant frequency $f_{21}(T)$ as the temperature increases, with a scaling factor of 2.44 GHz/volt.

To convert this to a frequency shift from zero temperature, $\Delta f_{21}(T) = f_{21}(T) - f_{21}(0)$, we require a zero-temperature reference voltage $V_{21}(0)$ which would be the resonant position at $T = 0$. For each of the two long experimental runs the data was extrapolated to zero temperature by fitting to a power law $\Delta f_{21}(T) = -AT^{\alpha}$. An excellent fit over the whole temperature range was found with A = 810 ± 100 MHz and α = 2.5 ± 0.2. In practice the reference voltage was very constrained, leading to an uncertainty of less than ± 3 MHz at 0.1 K, which is very small compared to 189.6 GHz and to the 810 MHz shift at 1 K. The resonant frequency decreased by 0.43% at 1 K. The data is plotted in Fig. 7.

The measured shift $\Delta f_{21}(T)$ is independent of electron density as shown for experimental runs with electron densities from 0.67 to 2.4 × $10^{11}$ m$^{-2}$. Some spot points were also taken at densities up to 6.5 × $10^{11}$ m$^{-2}$. It is also independent of the microwave power in the low-power limit and of the liquid helium level in the cell (but see section E below).

The resonant frequency shift $\Delta f_{21}(T)$ is due to ripplons on the helium surface but the theory is subtle, challenging and interesting as shown recently by Konstantinov *et al*. [22]. The electron energy spectrum is renormalised by interactions with zero-point and thermal ripplons. In particular, two-ripplon processes result in non-zero diagonal matrix elements of the interaction Hamiltonian and are responsible for temperature dependent Lamb-like shifts of the eigenvalue energies. The final result for the theoretical temperature dependent frequency shift $\Delta f_{21}(T)$ for $E_z$ = 106 V/cm is plotted in Fig. 7. Given that there are no adjustable parameters in the calculations, this agreement is good, particularly below 1 K..

### E. Density dependent resonant frequency

The temperature dependent frequency shift was independent of the electron density. But, for a given pressing voltage $V_z$, the electric field $E_z$ and hence the resonant frequency will depend on the electron density through Eq. (8c), if the helium level is not precisely at the center of the capacitor plates [2], giving a voltage offset and an apparent density dependent frequency shift. In the two runs where this was investigated, the resonance position in $V_z$ and the equivalent frequency shift $\Delta f_{21}$, varied linearly with the electron density $n$, as shown in Fig. 8. This corresponds to the liquid helium surface being 38 μm above and 3 μm below the center of the cell for the two runs respectively. This gives an additional electric field $E_z$ proportional to $n$, from the second term in Eq. (8c). The measured temperature-dependent

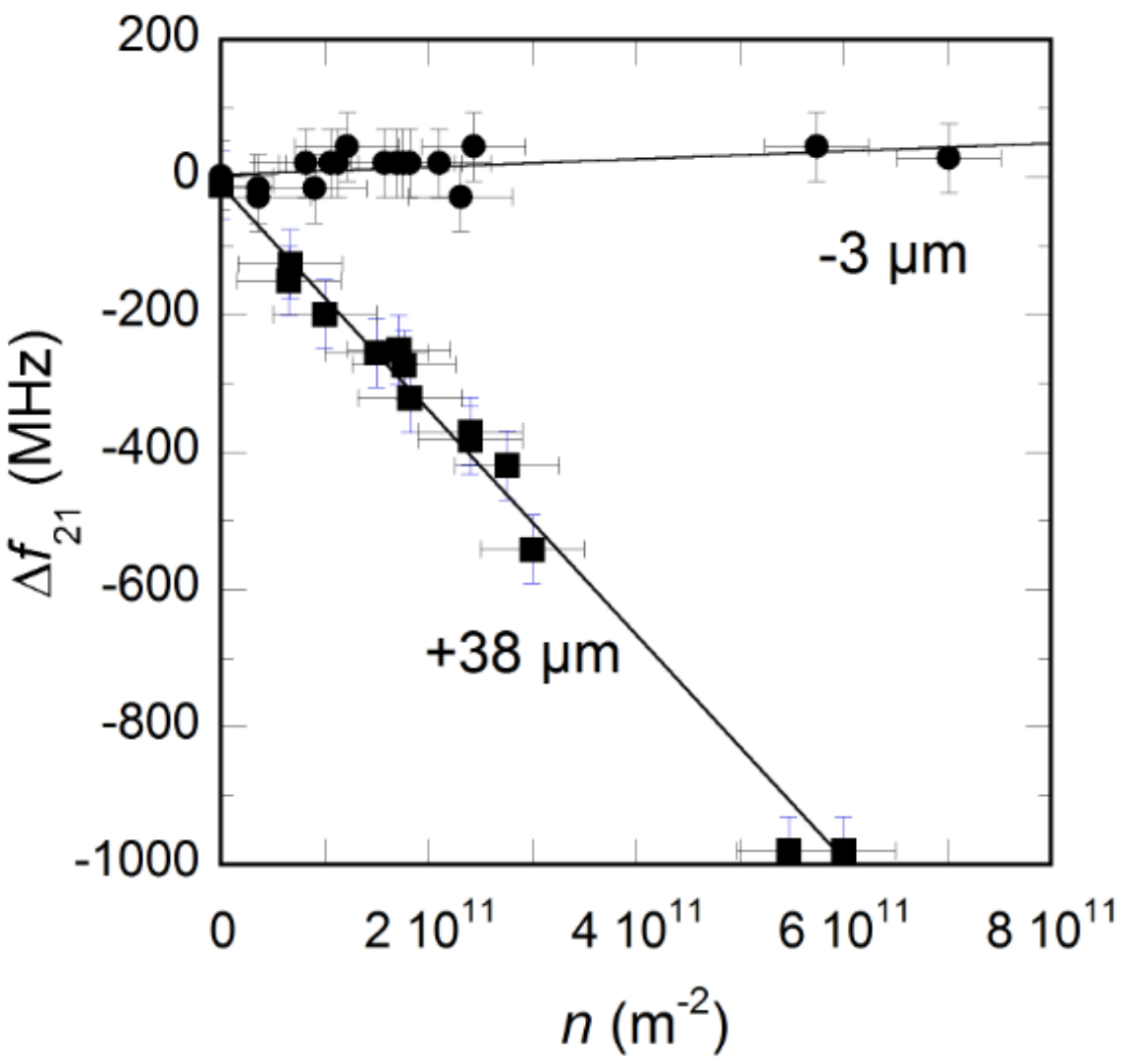

FIG. 8. The change in resonant frequency, extrapolated to zero temperature, versus the electron density $n$ for two separate experimental runs. The lines show the predicted density dependence if the helium level lies 3 µm below (●), and 38 µm above (■), the center of the cell capacitor respectively.

frequency shifts were the same in both cases, within the error bars.

This enables us to eliminate another possible cause of a resonant frequency shift with temperature, from changes in the helium level. The surface tension decreases with temperature and could change the helium level, through capillary rise, though the change is small, 2% at 1 K [26]. The level will also fall due to an increasing helium vapor density above 1 K. This would give a temperature-dependent field $E_z$, through Eq. (8c), and a temperature dependent frequency shift. But this would be density dependent, particularly at higher temperatures, which was not observed. Using Eq. (8), we estimate that the frequency shift for a height change would be 1.5 MHz/µm for no electrons; -2.8 MHz/µm for $n = 1 \times 10^{11}$ m$^{-2}$; and -8.8 MHz/µm for $n = 2.4 \times 10^{11}$ m$^{-2}$. We conclude that any changes in the helium height did not contribute to the temperature-dependent frequency shifts, within the error bars.

Some measurements were made in the Wigner crystal, below the melting temperature $T_m = 0.225 \times 10^{-6}\, n^{1/2}$ K[27]. For the highest density used here $n = 6.7 \times 10^{11}$ m$^{-2}$, $T_m = 0.181$ K. No effects were observed but no systematic measurements were made.

## IV. CONCLUSION

We have presented measurements of the resonant microwave absorption by the Rydberg energy levels of surface state electrons on superfluid helium, in the frequency range 165 – 220 GHz. The differential absorption was measured and integrated to give the microwave absorption lineshape. Direct measurements of the absorption line were made using large-amplitude square wave-modulation.

The microwave lineshape was measured at low microwave powers from 0.1 to 1.5 K. The small temperature independent residual lineshape at low temperatures was probably due to inhomogeneous broadening. A convolution analysis was used to determine the temperature-dependent Lorentzian contribution to the linewidth $\gamma(T)$ for excitation to the first excited state. In the ripplon scattering region below 0.7 K, the experimental results are consistent with the theory by Ando [13]. In the gas atom scattering regime, the linewidth is proportional to the vapor atom number density, as expected, but with a scaling factor of 1.6 above Ando's numerical result, confirming a previous analysis by Isshiki *et al*. [7].

The resonant frequency was found to be temperature dependent, decreasing as temperature increases. Below 1 K this is in good agreement with a recent theory by Konstantinov *et al*. [22] based on the renormalization of the electron energy spectrum due to dynamic interactions with ripplon excitations, analogous to a Lamb shift in the electronic energy levels.

## ACKNOWLEDGEMENTS

We thank A. J. Dahm, J. Goodkind, K. Kono, S. A. Lyon, P. J. Meeson, Y. Mukharsky, P. M. Platzman, J. Saunders and especially M. I. Dykman and D. Konstantinov for discussions and F. Greenough, A. K. Betts and others for technical support. The work was supported by the EPSRC, by the EU Human Potential Programme under contract HPRN-CT-2000-00157 *Surface Electrons*, and by Royal Holloway, University of London.

# SUPPLEMENTAL MATERIAL
# Temperature dependent energy levels of electrons on liquid helium

This Supplemental Material describes the technical details of the microwave equipment and cell used in these experiments, including the microwave source, the waveguide and components, the microwave cell and the InSb, or Putley, detector.

### I. Microwave source

The microwave system used is shown schematically in Fig. 9. Microwave power was generated by a Gunn diode oscillator [28] (82.5 – 97.5GHz, minimum output 30 mW) and passed through a doubler (5 mW maximum output from 165 to 195 GHz) and transmitted down overmoded waveguide, through thermal filters, into the experimental cell mounted on a dilution refrigerator. The frequency of the Gunn oscillator was phase-locked to a 10 MHz quartz crystal resonator. Higher frequencies, up to 260 GHz, were obtained from a Carcinotron source. The most detailed experiments were done at a frequency of 189.6 GHz.

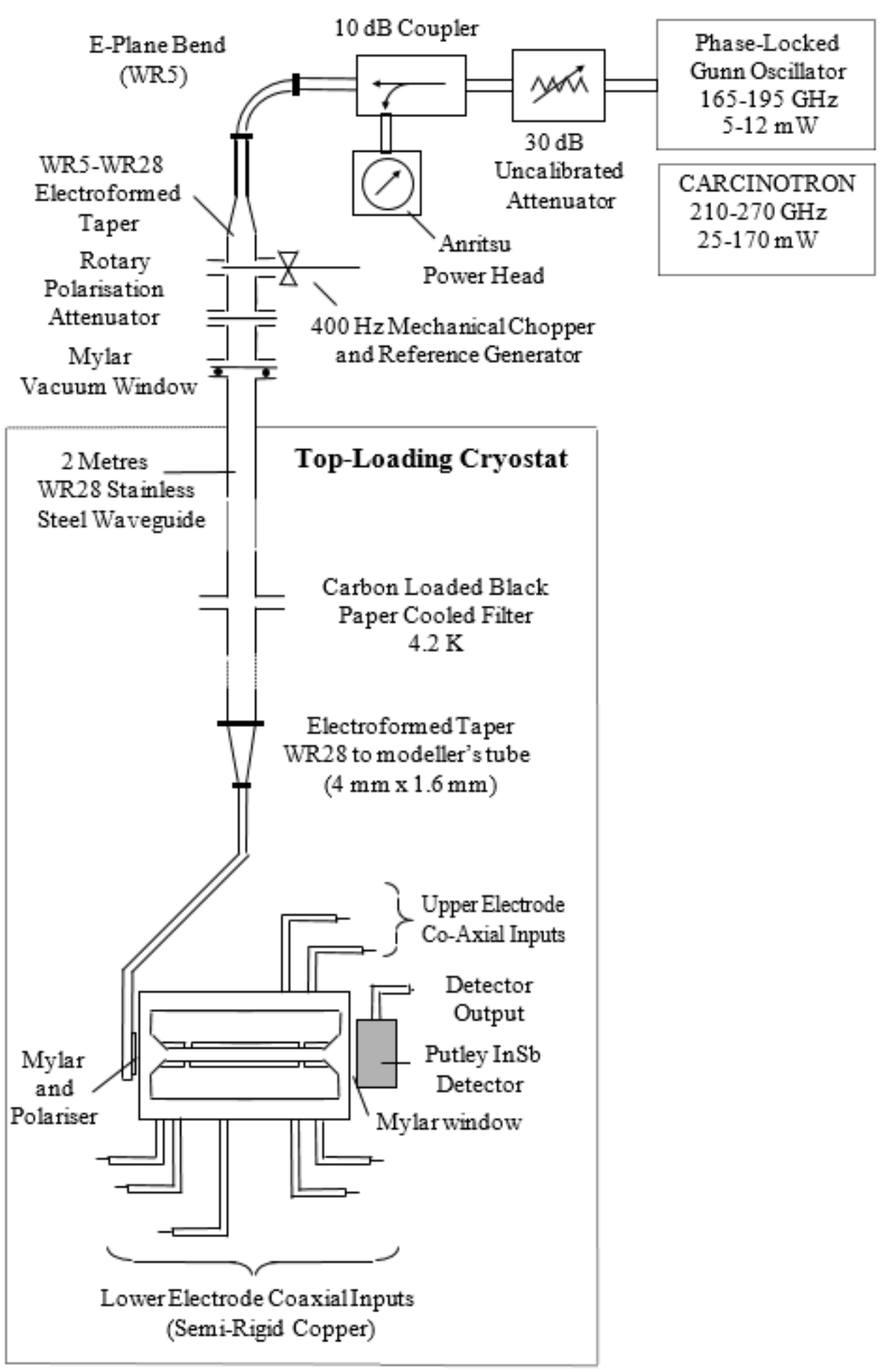

FIG. 9. Schematic diagram of the microwave system used.

### II. Microwave cell

The cell is shown schematically in Fig. 1. The electrons were held between two circular upper and lower capacitor plates, radius 28 mm, with an electrode separation of $h$ = 2.075 mm. These were fabricated on microwave micro-circuit board with polished OFHC copper, plated with a 1 µm flash of gold. The lower electrode had a central disc of radius 22 mm, and an outer annular ring with 4 equal segments, separated by a 0.5 mm gap of dielectric as shown in Fig 4. The electrodes were mounted on copper plates in the cell which acted as a guard.

The free electrons were produced by thermionic emission from a pulsed filament in the upper cell..

The electron magnetoconductivity was measured using a Corbino geometry to obtain the electron density and mobility. The central electrode was driven with a 10 kHz ac signal which was detected on the lower outer ring electrodes via capacitive coupling to the electrons. The currents to the four electrode segments were balanced to level the cell with the electrons in place. The electron density profile was calculated to be constant out to a radius of about 27 mm, depending on the potentials.

The two electrodes form a microwave pillbox cavity. When the frequency is swept, various resonant modes are excited. For a maximum electric field amplitude $E_z$ on the electrons, we require a $TM_{mnp}$ mode with $p$ = 2, where the mode numbers refer to the azimuthal, radial and axial directions. The mode number $p$ is the number of standing half-wavelengths along the $z$-axis. For this mode the $z$-axis component of the microwave field is a maximum in the middle of the cell where the electrons are distributed. For our cell, the low frequency cut-off for these modes would be 150 GHz. Experiments on a test cell clearly showed these volume-mode resonances and the onset of the $p$ = 2 modes. Above the cut-off frequency, the cell is overmoded with a wide range of $m$ and $n$ resonances, and was tuned to maximise the signal output. Once tuned, it was very stable. The microwave intensity may have nodes and anti-nodes across the electron sheet, with a typical node spacing of about 1 mm, much smaller than the sheet diameter.

The input and output microwave ports were vacuum-sealed using Mylar windows and indium O-rings. The microwaves were polarized vertically by a horizontal wire grid (20 lines/mm) on the input port and propagated horizontally through the cell to the output port, which was coupled to an InSb hot-electron, or Putley [29], bolometer [30] in a carefully designed reflective housing. The detector element was thermally isolated by a thin (0.5 mm) PTFE gasket. A d.c. bias current $I$, up to 50 μA, was passed through the detector from a voltage source, using a current-limiting resistance of $10^5$ Ω. The resistance $R(P)$ of the detector element was both temperature and power dependent, where the power $P$ dissipated comes from the Joule heating $P = I^2R(P)$ and from the absorbed microwave power. The resistance at low power increased from 5 kΩ at 4.2 K to 26.5 kΩ at 0.9 K and to 160 kΩ at 0.5 K. The voltage sensitivity $S = -\, dV/dP = -\, I\, dR(P)/dP$ of the detector was calculated by measuring $R(P)$ as a function of bias current. The detector sensitivity was a maximum for a bias voltage of about 300 mV, corresponding to a bias current of 20 μA. The voltage output from the detector was amplified by a factor of 100 using a preamplifier. The overall maximum sensitivity, measured at the preamplifier output, is about 2.4 V/μW at 0.9 K, compared to a specified sensitivity of 0.6 V/μW at 4.2 K. The sensitivity increased further at lower temperatures, depending on the bias current, which had to be reduced at lower temperatures to reduce Joule heating. The noise-equivalent power (NEP) of the Putley detector was typically 0.5 pW/√Hz with a preamplifier input noise of 1.1 nV/√Hz.

The microwave power output of the oscillator was measured at room temperature with an Anritsu Power Meter ML83A. The power passing through the cell was measured with the Putley detector. The microwaves were modulated using several techniques and the modulation was measured with a lock-in amplifier following the preamplifier. With no electrons on the helium, a mechanical 6-bladed chopper was installed in the waveguide at room temperature. A photodiode was used to monitor the rotation of the chopper and to synchronize the lock-in amplifier (at about 330 Hz). The output from the Putley detector for the chopped microwaves was used (i) to monitor the microwave frequency dependent transmission through the cell and (ii) to measure the microwave input power during the absorption experiments before the helium was charged. The linearity of the Putley detector was carefully checked at all operating temperatures and was accurately linear up to an output voltage greater than 1000 mV, corresponding to some 400 nW absorbed in the InSb element. The transmission loss from the source through to the detector was typically in the range −23 to −40 dB.

---

[28] Radiometer Physics Gmbh.
[29] E. H. Putley, Appl. Opt. **4**, 649 (1965).
[30] QMC Instruments Ltd.